\documentclass[conference]{IEEEtran}
\IEEEoverridecommandlockouts
\usepackage{amsmath,amssymb,amsfonts}

\usepackage{algpseudocode}
\usepackage{graphicx}
\usepackage{textcomp}
\usepackage{booktabs}
\usepackage{enumitem}
\makeatletter
\let\c@lofdepth\relax
\let\c@lotdepth\relax
\makeatother
\usepackage[caption=false]{subfig}
\usepackage{caption}

\usepackage[colorlinks,bookmarksopen,bookmarksnumbered,citecolor=red,urlcolor=red]{hyperref}

\def\BibTeX{{\rm B\kern-.05em{\sc i\kern-.025em b}\kern-.08em
T\kern-.1667em\lower.7ex\hbox{E}\kern-.125emX}}

{ \bgroup
  \addtolength\abovedisplayshortskip{#1}
  \addtolength\abovedisplayskip{#1}
  \addtolength\belowdisplayshortskip{#1}
  \addtolength\belowdisplayskip{#1}}
{\egroup\ignorespacesafterend}

\begin{document}

\newcommand{\tabincell}[2]{\begin{tabular}{@{}#1@{}}#2\end{tabular}}


%

%

%
\ifCLASSINFOpdf
\else
\fi
\hyphenation{op-tical net-works semi-conduc-tor}

\newenvironment{figurehere}
{\def\@captype{figure}}
{}
\makeatother

\title{QoE-driven Coupled Uplink and Downlink Rate Adaptation for 360-degree Video Live Streaming}


\author{Jie Li,
      Ransheng Feng,
      Zhi Liu,~\IEEEmembership{Senior Member,~IEEE,}
      Wei Sun,
      Qiyue Li,~\IEEEmembership{Member,~IEEE}

\thanks{Manuscript received August 28, 2019; revised December 10, 2019; accepted
January 03, 2020. This research is supported in part by National Natural Science Foundation of China, Grant No. 51877060, the Fundamental Research Funds for the Central Universities, Grant No. JZ2019HGTB0089 and PA2019GDQT0006, the State Grid Science and Technology Project (Research and application of key Technologies for integrated substation intelligent operation and maintenance based on the fusion of heterogeneous network and heterogeneous data). The work of Zhi Liu was supported by JSPS KAKENHI Grants 18K18036, 19H04092, and the Telecommunications Advancement Foundation Research Fund. (Corresponding author: Qiyue Li.)}

\thanks{Jie Li and Ransheng Feng are with School of Computer and Information, Hefei University of Technology, Hefei, 23009, China. e-mail: lijie@hfut.edu.cn, ranshengfeng@mail.hfut.edu.cn}

\thanks{Zhi Liu is with Department of Mathematical and Systems Engineering, Shizuoka University, Japan. e-mail: liu@ieee.org}

\thanks{Wei Sun and Qiyue Li are with School of Electrical Engineering and Automation, Hefei University of Technology Hefei, 23009, China. e-mail:wsun@hfut.edu.cn, liqiyue@mail.ustc.edu.cn}

\thanks{\textcircled{c} 20XX IEEE.  Personal use of this material is permitted.  Permission from IEEE must be obtained for all other uses, in any current or future media, including reprinting/republishing this material for advertising or promotional purposes, creating new collective works, for resale or redistribution to servers or lists, or reuse of any copyrighted component of this work in other works.}

}



\maketitle

\begin{abstract}
360-degree video provides an immersive 360-degree viewing experience and has been widely used in many areas. The 360-degree video live streaming systems involve capturing, compression, uplink (camera to video server) and downlink (video server to user) transmissions. However, few studies have jointly investigated such complex systems, especially the rate adaptation for the coupled uplink and downlink in the 360-degree video streaming under limited bandwidth constraints. In this letter, we propose a quality of experience (QoE)-driven 360-degree video live streaming system, in which a video server performs rate adaptation based on the uplink and downlink bandwidths and information concerning each user's real-time field-of-view (FOV). We formulate it as a nonlinear integer programming problem and propose an algorithm, which combines the Karush-Kuhn-Tucker (KKT) condition and branch and bound method, to solve it. The numerical results show that the proposed optimization model can improve users' QoE significantly in comparison with other baseline schemes.
\end{abstract}


%

\section{Introduction} \label{sec_intro}

360-degree video technology has recently become more and more popular with the increasing demands on interactive applications.
By wearing a head-mounted display (HMD), 360-degree video users can freely move their heads to change the viewing directions, which provides an immersive viewing experience. To improve the quality of experience (QoE) \cite{8647729}, most 360-degree videos have 6K or even higher resolution. Streaming such high-resolution videos is non-trivial because of the limited bandwidth of wireless communication channels \cite{ liu2018jet}.
In addition, there has been a recent trend toward high-quality 360-degree video content creation using 3D panoramic VR cameras. Compared to the traditional live streaming, 360-degree video live streaming is considerably more challenging due to its panoramic nature \cite{Liu2019},
and it has more stringent QoE requirements to prevent motion sickness.



Many published works have investigated 360-degree video streaming \cite{3123372,2017,3123291,Konrad2017,guo2018optimal}. For example, the authors of \cite{2017} designed a rate adaptation algorithm that can maximize the defined QoE metrics for 360-degree video streaming. Given the FOV and bandwidth estimation, Xie et al. \cite{3123291} proposed a probabilistic tile-based adaptive 360-degree video streaming system, named \textit{360ProbDASH}, which combined viewport adaptation and rate adaptation to solve the QoE-driven optimization problem. In \cite{Konrad2017}, the authors presented \textit{Vortex}, a live VR video streaming system that works in a computationally and bandwidth-efficient manner.

In 360-degree video live streaming applications, a capturing device is used to record scene and transmit the captured video to the server in real time. Then, multiple users can access and watch the 360-degree video with their HMDs by downloading the coresponding video content from the server. Obviously, under a limited bandwidth constraint, the uplink (VR cameras to video server)  and downlink rate (video server to end users) should be carefully selected to improve the viewing experiences. 

In this paper, we formulate this transmission problem as a nonlinear integer programming problem. To solve this, we propose an optimal algorithm that combines the KKT condition and the branch and bound method. Extensive simulations are conducted based on the real-world LTE network traces, and the simulation results show that the proposed solution can significantly improve users' QoE in  comparison with other baseline schemes. To the best of our knowledge, this is the first research jointly considering rate adaptation for coupled uplink and downlink in 360-degree video streaming system.



\section{System Overview} \label{sec_system_model}

In this section, we present our 360-degree video live streaming system, which jointly allocates uplink and downlink wireless resources to maximize the overall QoE.

\subsection{System Model} \label{sec_wireless_transmission_model}

\begin{figure}[tb]
\setlength{\belowcaptionskip}{4mm} 
	\centering
	\includegraphics[width=3in]{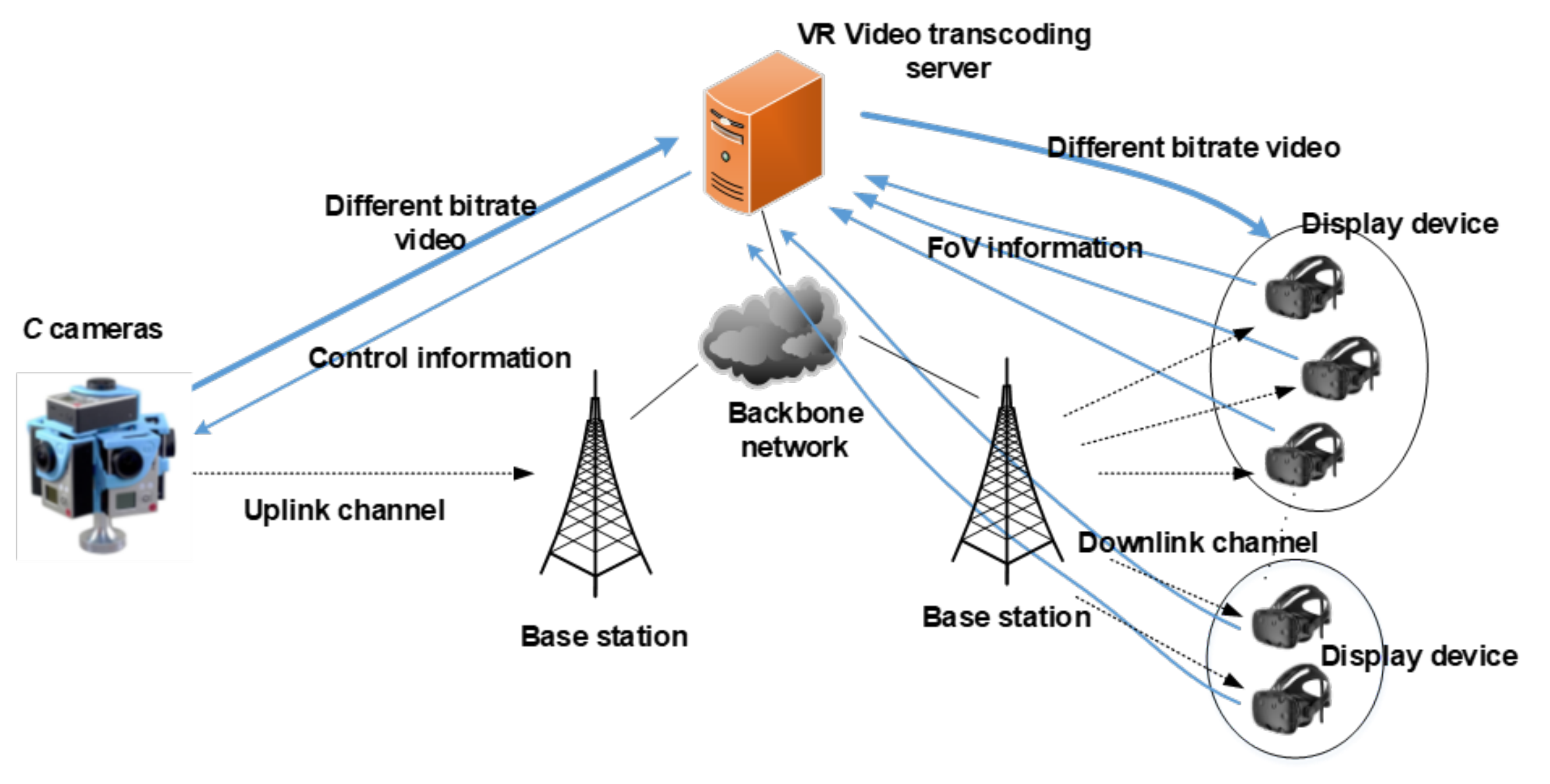}
\caption{A typical scenario of a 360-degree video live streaming system.}
	\label{fig_application_scenario}
\end{figure}

We consider a scenario where multiple users are watching 360-degree video live streaming, as illustrated in Fig. \ref{fig_application_scenario}. $C$ cameras are installed in the 360-degree video capturing device, and each camera can record high-quality 2D video. Constrained by the limited wireless uplink bandwidth, the recorded $C$ videos may not be transmitted to video server with the highest quality. Assuming that each camera is capable of encoding the original video into several versions with different bitrates, the video with the most appropriate bitrate should be selected and uploaded to the video server.
Video is partitioned into tiles at the server and the corresponding tiles for user's FoV are transmitted to the user from the server. The downlink transmission is also limited by the bandwidth. Then, one representation is selected for every tile and transmitted to the user, according to the corresponding FOV and channel status under the assumption that such information is transmitted to the video server in real time. Obviously, for each tile, the best quality received by a user cannot exceed the quality of the video uploaded by the corresponding camera, where the original video is captured. Thus, this scenario can be modeled as a typical coupled video transmission problem.

\begin{figure}[tb]
\setlength{\belowcaptionskip}{0mm} 
	\centering
	\includegraphics[width=2.8in]{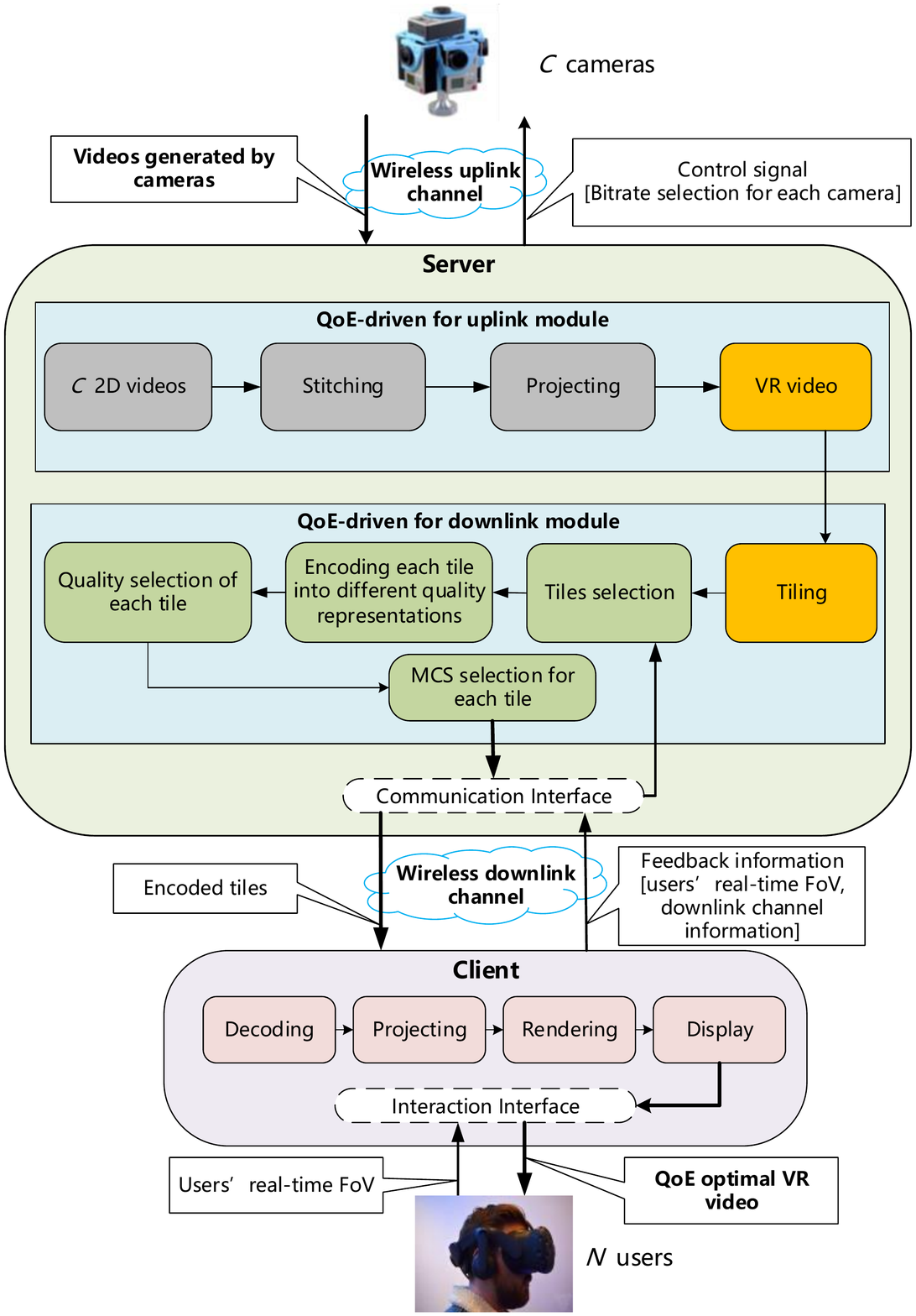}
\caption{QoE driven coupled uplink and downlink 360-degree video transmission system.}
	\label{fig_system_arch}
\end{figure}

In this paper, we consider this coupled uplink and downlink transmission system to maximize the QoE of all users through rate adaptation. The architecture is shown in Fig. \ref{fig_system_arch}.  The system consists of four parts: a 360-degree video capturing device with $C$ cameras, processing modules on both the server side and the client, and the end users wearing HMDs.
Each camera can encode the captured video into different video qualities, expressed as different constant bitrates, but only one video bitrate can be uploaded to the video server.

The server side includes two modules: the QoE-driven uplink processing module and the downlink processing module. The uplink processing module is responsible for obtaining a raw 360-degree video by stitching the $C$ uploaded videos. The downlink processing module selects the appropriate video quality for each tile based on the FOVs and channel information from all the users. The system workflow is as follows: the uplink processing module selects a bitrate level for each camera. Then, the $C$ videos with different qualities are stitched and projected into a panoramic video. Next, the downlink processing module divides the raw panoramic video into tiles and encodes each tile into different quality representations. With the help of feedback information from the client side, the downlink processing module can determine the range of tiles to be transmitted and select the appropriate quality level for each tile. Finally, all the selected tiles are transmitted through the downlink channel to maximize the users' QoE.

At the client side, as the tiles are received from the server, they are decoded, projected, rendered and displayed in the HMD. In addition, the client side sends the user's FOV and channel information in real time to the video server through each user's uplink feedback channel.

\vspace{-0.1em}
\subsection{Problem Formulation} \label{sec_problem_formulation}
Suppose the system includes $N$ users (indexed by $n$). The evenly deployed $C$ cameras form a 360-degree video capturing system. Each camera $c$ can record and generate videos at up to ${D}^{'}$ bitrate levels (indexed by ${d}^{'}$). The bitrate of camera $c$ uploaded with bitrate level ${d}^{'}$ is denoted by $\lambda_{c,d^{'}}^{UL}$. The total bandwidth of the uplink channel is $B^{UL}$. When the videos generated by the $C$ cameras are uploaded to the server, the server processes them to produce a 360-degree video. Prior to downlink transmission, the 360-degree video is divided into $T$ tiles (indexed by $t$), and each tile is encoded into $D$ different representations (indexed by $d$) at different quality levels, which is the same as how the HTTP DASH processes the video. We denote the bitrates of tile $t$ with representation $d$ in GOP $k$ by  $\lambda_{t,d,k}^{DL}$. As aforementioned, the quality of each video $d$ in the downlink cannot exceed the quality of the corresponding video ${d}^{'}$ in the uplink transmission. Assume that the bandwidth of the k-th GOP in the downlink channel is $B_{k}^{D L}$. For user $n$, $T_{FoV}^{n}$ denotes the tiles covered by his FOV. Then user $n$'s expected QoE can be defined as follows:
\begin{equation}
\begin{small}
\begin{array}{l}
QoE_{n} = \sum\limits_{\forall t \in T_{FOV}^n} {\sum\limits_{\forall d \in D} {\sum\limits_{\forall k \in K} {q\left( {\lambda _{t,d,k}^{DL} \cdot \chi _{t,d,k}^{DL}} \right)} } }  - \alpha \times\\
 \sum\limits_{\forall t \in T_{FOV}^n} {\sum\limits_{\forall d \in D} {\sum\limits_{\forall k \in K} {\mathbb {I} \left( {\lambda _{t,d,k}^{DL} \cdot \chi _{t,d,k}^{DL} > B_k^{DL}} \right)} } }  \cdot {T_k} - \beta \times\\
\sum\limits_{\forall t \in T_{FOV}^n} {\sum\limits_{\forall d \in D} {\sum\limits_{\forall k \in K} \left(q\left(\lambda_{t, d, k+1}^{D L} \cdot \chi_{t, d, k+1}^{D L}\right)-q\left(\lambda_{t, d, k}^{D L} \cdot \chi_{t, d, k}^{D L}\right)\right)^{2} }},
\end{array}
\end{small}
\end{equation}
where $\lambda _{t,d,k}^{DL}$ denotes the bitrate of tile $t$ with bitrate level $d$ in the $k$-th GOP, and $DL$ means ``downlink" channel. $\chi _{t,d,k}^{DL}$ is a binary variable, which equals 1 if tile $t$ is transmitted with bitrate level $d$ in the $k$-th GOP, and 0 otherwise. Function $q$ is a mapping function, which maps the bitrate of tile $t$ to the quality perceived by the user. The form of function $q$ is the logarithmic of the received bitrate. The second term of this equation is used to show the impact of stalls. We assume stall will occur when the bandwidth for the $k$-th GOP is less than the video bitrate, and stall time is approximately equal to the duration of the $k$-th GOP \footnote{In a practical system, due to the existence of the playback buffer, stalls are related to the transmission rate, the playback speed and buffer status. Once the buffer drains, a stall will occur. In this manuscript, to simplify the formulation, we assume when the bandwidth value in a GOP is less than the video bitrate, stall will occur. How to more previously model this is left as the future work.}. $\mathbb {I}\left(\lambda_{t, d, k}^{D L} \cdot \chi_{t, d, k}^{D L}>B_{k}^{D L}\right)$ is an indicator function, and its value is 1 only when $\lambda_{t, d, k}^{D L} \cdot \chi_{t, d, k}^{D L}>B_{k}^{D L}$, otherwise 0. This means that in one GOP, when the video bitrate is greater than the bandwidth, the stalls will occur. $T_{k}$ is the duration of the $k$-th GOP. $B_{k}^{D L}$ is the bandwidth when transmitting the $k$-th GOP. Thus we can use the average of the bandwidth during $T_{k}$ as the value of $B_{k}^{D L}$. In addition, $\left(q\left(\lambda_{t, d, k+1}^{D L} \cdot \chi_{t, d, k+1}^{D L}\right)-q\left(\lambda_{t, d, k}^{D L} \cdot \chi_{t, d, k}^{D L}\right)\right)^{2}$ considers the quality switches between the consequent GOPs. Similarly, $\lambda_{t, d, k+1}^{D L}$ denotes the bitrate of tile $t$ with bitrate level $d$ in the $k+1$-th GOP. $\chi_{t, d, k+1}^{D L}$ is also a binary variable, and equals 1 if tile $t$ is transmitted with bitrate level $d$ in the $k+1$-th GOP,  and 0 otherwise. Finally, constants $\alpha$ and $\beta$ are the non-negative weight parameters to balance the three factors.
With this QoE model, our uplink and downlink optimization problem is defined as follows:\\
problem 1:
{\setlength\abovedisplayskip{0.5pt}
 \setlength\belowdisplayskip{0.5pt}
\begin{equation}
\begin{small}
\max \sum_{\forall n \in N} QoE_{n}
\end{small}
\end{equation}
}
s.t.
{\setlength\abovedisplayskip{0.5pt}
 \setlength\belowdisplayskip{0.5pt}
\begin{equation}
\begin{small}
\sum_{\forall d^{\prime} \in D^{\prime}} \chi_{c, d^{\prime}}^{U L}=1, \forall c \in C
\end{small}
\end{equation}
}
{\setlength\abovedisplayskip{0.5pt}
 \setlength\belowdisplayskip{0.5pt}
\begin{equation}
\begin{small}
\sum_{\forall d^{\prime} \in D^{\prime}} \sum_{\forall c \in C} \lambda_{c, d^{\prime}}^{U L} \cdot \chi_{c, d^{\prime}}^{U L} \leq B^{U L}
\end{small}
\end{equation}
}
{\setlength\abovedisplayskip{0.5pt}
 \setlength\belowdisplayskip{0.5pt}
\begin{equation}
\begin{small}
\sum_{\forall d \in D} \chi_{t, d, k}^{D L}=1, \forall t \in T_{F O V}^{n}, \forall k \in K
\end{small}
\end{equation}
}
{\setlength\abovedisplayskip{0.5pt}
 \setlength\belowdisplayskip{0.5pt}
\begin{equation}
\begin{small}
\sum_{\forall t \in T_{F O V}^{n}} \sum_{\forall d \in D} \sum_{\forall k \in K} \lambda_{t, d, k}^{D L} \cdot \chi_{t, d, k}^{D L} \leq \sum_{\forall k \in K} B_{k}^{D L}
\end{small}
\end{equation}
}
{\setlength\abovedisplayskip{0.5pt}
 \setlength\belowdisplayskip{0.5pt}
\begin{equation}
\begin{split}
\sum_{\forall d \in D} \sum_{\forall k \in K} \lambda_{t, d, k}^{D L} \cdot \chi_{t, d, k}^{D L} \leq \frac{1}{T} \sum_{\forall d^{\prime} \in D^{\prime}} \lambda_{c, d^{\prime}}^{U I} \cdot \chi_{c, d^{\prime}}^{U L},\\
\forall c \in C, \forall t \in T_{F O V}^{n}
\end{split}
\end{equation}
}
where $\chi_{c, d^{\prime}}^{U L}$  and $\chi _{t,d,k}^{DL}$ are the optimization variables. $UL$ stands for "uplink". $\chi_{c, d^{\prime}}^{U L}$ is a binary variable, which equals 1 when the video from camera $c$ is transmitted with bitrate level $d^{\prime}$ and 0 otherwise. $\chi _{t,d,k}^{DL}$  is also binary variable (same as in equation (1)), which equals 1 if tile $t$ is transmitted with representation $d$ in $k$-th GOP and 0 otherwise. $\lambda_{c, d^{\prime}}^{U L}$  denotes the bitrate of camera $c$ uploaded at bitrate level $d'$. Constraints (3)-(4) apply to the uplink. Constraint (3) indicates that the video of camera $c$ can be uploaded with only one quality level. The total bitrate of all the uploaded videos cannot exceed the total bandwidth of the uplink as specified in Constraint (4). Constraints (5)-(6) are the downlink constraints. Constraint (5) ensures that only one representation can be selected for tile $t$ in k-th GOP, which is transmitted to the client side. Constraint (6) ensures that the sum of the tile bitrates cannot exceed the total bandwidth of the downlink channel. Constraint (7)  discusses the coupled uplink and downlink and ensures that the quality levels of tiles in the downlink cannot exceed the quality level of the videos generated by the corresponding camera and transmitted in the uplink.

\section{QoE driven Optimal Rate Adaptation Algorithm} \label{sec_optimal_algorithm}

In this section, we introduce the optimal solution of the above problem, which is a nonlinear integer programming problem and can be proven to be NP-hard.  We firstly approximate the indicator function in the QoE model with a logarithmic function, so that the QoE function becomes a continuous function. Then, because the constraint in problem 1 satisfies the linear constraint qualification, we can use the KKT condition to solve the relaxation problem of problem 1. By using the logarithmic function to approximate the indicator function in the QoE model and relaxing the integer variables $\chi_{c,d^{'}}^{UL}$ and $\chi_{t,d,m}^{DL}$ to continuous variables, the relaxed problem can be solved by applying the KKT conditions and the Lagrangian function. Then, we can obtain the optimal value of the original problem.

\subsection{KKT condition for the relaxed continuous problem} \label{sec_KKT_condition}

First, we approximate the indicator function in the QoE model with a logarithmic function and relax $\chi_{c,d^{'}}^{UL}$ and $\chi_{t,d,k}^{DL}$ to continuous variables. Then, the original problem 1  can be solved by KKT conditions. The Lagrangian function of the problem is as follows Eq. (\ref{equ_Lagrangian_func}):

{\setlength\abovedisplayskip{1pt}
 \setlength\belowdisplayskip{1pt}
\begin{equation} \label{equ_Lagrangian_func}
\begin{split}
L(\chi _{c,{d^{'}}}^{UL},\chi _{t,d,k}^{DL},\lambda ,\mu ) =  - \sum_{\forall n \in N} {Qo{E_n}}  + {\lambda _1}h(\chi _{c,{d^{'}}}^{DL})\\
+ {\lambda _2}{h}(\chi _{t,d,k}^{DL}) + {\mu _1}g(\chi _{c,{d^{'}}}^{DL})\\
+ {\mu _2}g(\chi _{t,d,k}^{DL}) + {\mu _3}g(\chi _{c,{d^{'}}}^{DL},\chi_{t,d,k}^{DL}),
\end{split}
\end{equation}
}
where 
{\setlength\abovedisplayskip{1pt}
 \setlength\belowdisplayskip{1pt}
\begin{equation}
\begin{split}
h(\chi _{c,{d^{'}}}^{DL}){\text{ = }}\sum_{\forall d^{\prime} \in D^{\prime}} {\chi _{c,{d^{'}}}^{UL} - 1} \
\end{split}
\end{equation}
}
{\setlength\abovedisplayskip{1pt}
 \setlength\belowdisplayskip{1pt}
 \begin{small}
\begin{equation}
\begin{split}
{h}(\chi _{t,d,k}^{DL}){\text{ = }}\sum_{\forall d \in D} {\chi_{t,d,k}^{DL} - 1}
\end{split}
\end{equation}
\end{small}
}
{\setlength\abovedisplayskip{1pt}
 \setlength\belowdisplayskip{1pt}
\begin{small}
\begin{equation}
\begin{split}
g(\chi _{c,{d^{'}}}^{DL}){\text{ = }}\sum_{\forall d^{\prime} \in D^{\prime}} \sum_{\forall c \in C} {\chi _{c,{d^{'}}}^{UL} \cdot \lambda _{c,{d^{'}}}^{UL} - B{UL}} 
\end{split}
\end{equation}
\end{small}
}
{\setlength\abovedisplayskip{1pt}
 \setlength\belowdisplayskip{1pt}
\begin{equation}
\begin{split}
g(\chi _{t,d,k}^{DL}) = \sum_{\forall t \in T_{F O V}^{n}} \sum_{\forall d \in D} \sum_{\forall k \in K} \lambda_{t, d, k}^{D L} \cdot \chi_{t, d, k}^{D L} - \sum_{\forall k \in K} B_{k}^{D L}
\end{split}
\end{equation}
}
{\setlength\abovedisplayskip{1pt}
 \setlength\belowdisplayskip{1pt}
\begin{equation}
\begin{split}
g(\chi _{c,{d^{'}}}^{DL},\chi_{t,d,k}^{DL})=\sum_{\forall d \in D} \sum_{\forall k \in K} \lambda_{t, d, k}^{D L} \cdot \chi_{t, d, k}^{D L} \\
- \frac{1}{T} \sum_{\forall d^{\prime} \in D^{\prime}} \lambda_{c, d^{\prime}}^{U I} \cdot \chi_{c, d^{\prime}}^{U L}
\end{split}
\end{equation}
}
Thus, we can obtain the relevant KKT conditions:

{\setlength\abovedisplayskip{1pt}
 \setlength\belowdisplayskip{1pt}
\begin{equation}
\begin{split}
\frac{{\partial L(\chi _{c,{d^{'}}}^{UL},\chi _{t,d,k}^{DL},\lambda ,\mu )}}{{\partial \chi _{c,{d^{'}}}^{UL}}}{\text{ = }}{\lambda _1}\frac{{\partial h(\chi _{c,{d^{'}}}^{DL})}}{{\partial \chi _{c,{d^{'}}}^{UL}}}{\text{ + }}{\mu _1}\\
{\text{ + }}{\mu _3}\frac{{\partial g(\chi _{c,{d^{'}}}^{DL},\chi _{t,d,k}^{DL})}}{{\partial \chi _{c,{d^{'}}}^{UL}}}{\text{ = }}0
\end{split}
\end{equation}
}
{\setlength\abovedisplayskip{1pt}
 \setlength\belowdisplayskip{1pt}
\begin{equation}
\begin{split}
\frac{{\partial L(\chi _{c,{d^{'}}}^{UL},\chi _{t,d,k}^{DL},\lambda ,\mu )}}{{\partial \chi _{t,d,k}^{DL}}}{\rm{ = }} - \sum_{\forall n \in N} {\frac{{\partial Qo{E_n}}}{{\partial \chi _{t,d,k}^{DL}}}} {\rm{ + }}{\lambda _2}\frac{{\partial {h_1}(\chi _{t,d,k}^{DL})}}{{\partial \chi _{t,d,k}^{DL}}}\\
{\rm{ + }}{\mu _2}\frac{{\partial g(\chi _{t,d,k}^{DL})}}{{\partial \chi _{t,d,k}^{DL}}}{\rm{ + }}{\mu _3}\frac{{\partial g(\chi _{c,{d^{'}}}^{DL},\chi_{t,d,k}^{DL})}}{{\partial \chi_{t,d,k}^{DL}}}{\rm{ = }}0
\end{split}
\end{equation}
}
{\setlength\abovedisplayskip{1pt}
 \setlength\belowdisplayskip{1pt}
\begin{small}
\begin{equation}
\begin{split}
g(\chi _{c,{d^{'}}}^{DL}) \leqslant 0,g(\chi _{t,d,k}^{DL}) \leqslant 0,g(\chi _{c,{d^{'}}}^{DL},\chi _{t,d,k}^{DL}) \leqslant 0
\end{split}
\end{equation}
\end{small}
}
{\setlength\abovedisplayskip{1pt}
 \setlength\belowdisplayskip{1pt}
\begin{small}
\begin{equation}
h(\chi _{c,{d^{'}}}^{DL}){\text{ = 0,}}{\kern 1pt} {h}(\chi _{t,d,k}^{DL}){\text{ = }}0
\end{equation}
\end{small}
}
{\setlength\abovedisplayskip{1pt}
 \setlength\belowdisplayskip{1pt}
\begin{equation}
\begin{small}
{\lambda _1},{\lambda _2} \ne 0,{\mu _1},{\mu _2},{\mu _3} \geqslant 0
\end{small}
\end{equation}
}
{\setlength\abovedisplayskip{1pt}
 \setlength\belowdisplayskip{1pt}
\begin{equation}
\begin{small}
{\mu _1}g(\chi_{c,{d^{'}}}^{DL}){\text{ = }}0,{\mu _2}g(\varphi _{t,d,k}^{DL}) = 0,{\mu _3}g(\chi_{c,{d^{'}}}^{DL},\chi _{t,d,k}^{DL}) = 0.
\end{small}
\end{equation}
}
By solving equations (14)-(19), which are associated with the KKT condition, we can derive the optimal solution of the relaxed nonlinear problem. Next, we use the branch and bound method to find the solution of the original binary programming problem.

\subsection{Branch and bound Method} \label{sec_Branch-and-bound_method}

The branch and bound method designed to solve problem 1, as shown in Table \ref{tab:al}. The initial inputs are $\chi_{relax}$ and $Z_{relax}$, where $\chi_{relax}$ is the solution to the corresponding relaxation problem solved by the KKT condition, and $Z_{relax}$ indicates the corresponding optimal objective function value. The outputs are the 0--1 variable solution $\chi_{0-1}$ and the corresponding optimal objective function value $Z_{0-1}$.

\begin{table}[h]
	\centering
	\caption{Branch and Bound Algorithm for Solving Problem 1.}
	\begin{tabular}{l}\hline
	     \textbf{Input:} The optimal solution of the relaxation problem $\chi_{relax}$,\\\
		the optimal objective function value of the relaxation problem $Z_{relax}$, \\\
		and a random value in the range (0,1) $\epsilon$.\\\
		\textbf {Output:} The optimal solution of problem 1 $\chi_{0-1}$, \\\
		and the optimal objective function value of problem 1 $Z_{0-1}$.\\\
		\textbf{Initial:} $k=0$, $L=0$, $U=Z_{relax}$\\\
		 Choose any solution $\chi_{j}$ that does not meet the 0--1 constraints from $\chi_{relax}$, \\\
		 $\chi_{j} \in(0,1)$.\\\
		 \textbf {IF} {$0 < \chi_{j} < \epsilon$}\\\
		 Add the constraint $\chi_{j}=0$ to Problem P-1 to form subproblem I.\\\
		 \textbf {ELSE}\\\
          Add the constraint $\chi_{j}=1$ to Problem P-1 to form subproblem II.\\\
          \textbf {END IF}\\\
		 k++,find the solutions to the relaxation problems in subproblems I and II \\\
		 (denoted as $\chi_{k}$) where the optimal objective function value is $Z_{k}$ .\\\
		 Find the maximum value of the optimal objective function and use it as a \\\
		 new upper bound. Update  $L=max\{Z_{k}\}$, $\chi_{k}\in\{0,1\}$\\\
		 Then, find the maximum value of the objective function from the branch that meets \\\
		 the 0-1 condition as a new lower bound, and update $L=max\{Z_{k}\}$, $\chi_{k}\in\{0,1\}$.\\\
		  \textbf {IF} {$Z_{k}<L$}\\\
	     Cut off the bound \\\
		 \textbf {ELSE IF} {$Z_{k}>L$ and $\chi_{k} \in \{0,1\}$}\\\
		 Go to step 2\\\
		\textbf {ELSE} The optimal solution of problem 1 has been found, $\chi_{0-1}=\chi_{k}$ \\\
		and $Z_{0-1}=Z_{k}$\\\hline
	\end{tabular}
	\label{tab:al}
\end{table}

\section{Performance Evaluation} \label{sec_experiment_result}
We conduct experiments to verify the performance of the proposed QoE-driven 360-degree video live streaming system.

\subsection{Simulation Setup} \label{sec_simulation_setup}
\begin{figure}
	\begin{minipage}[tb]{0.44\linewidth}
		\centering
		\includegraphics[width=1.4in]{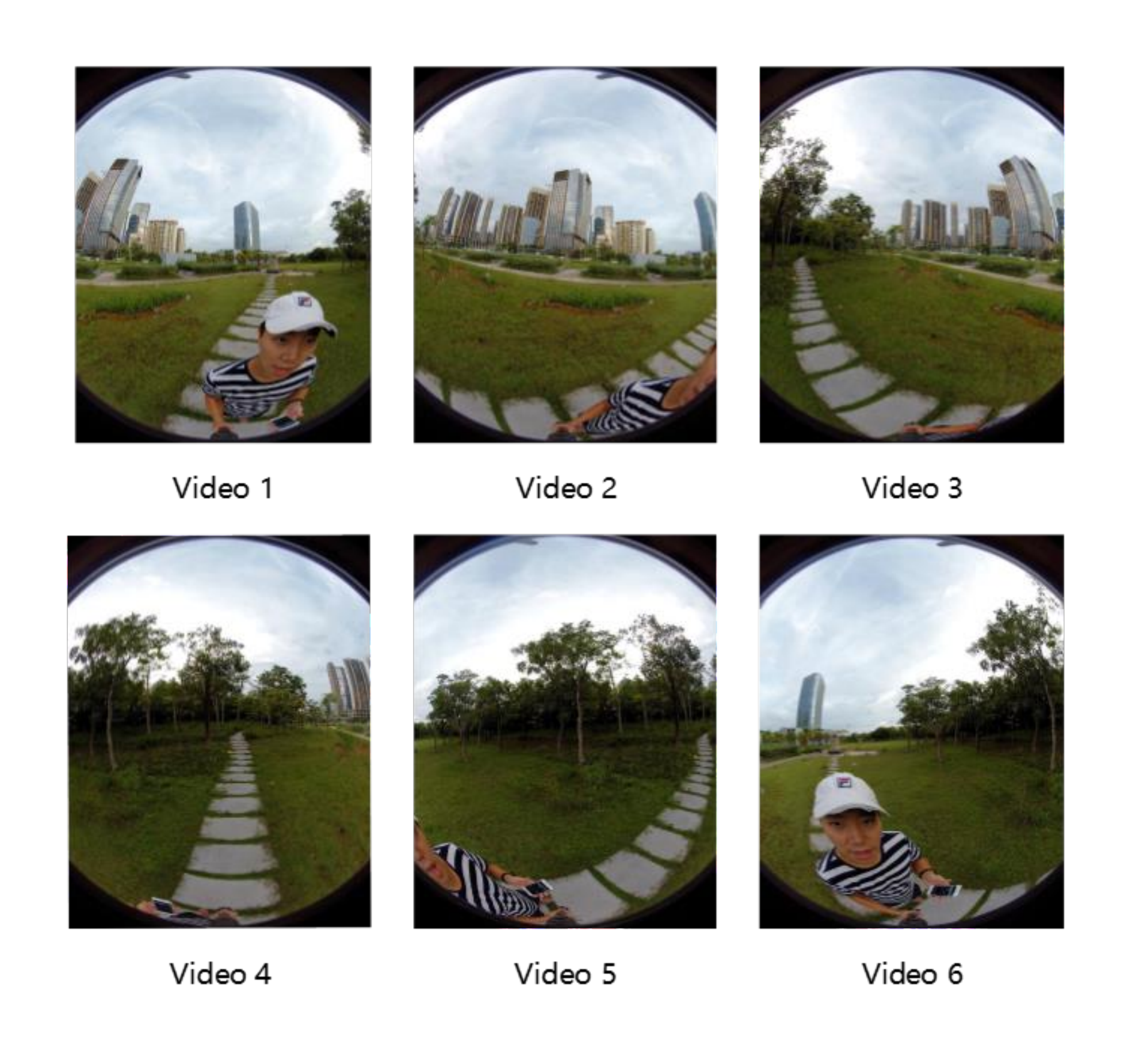}
		\caption*{(a) Snapshot of the six original videos.}
		\label{fig_snapshot_original_videos}
	\end{minipage}%
	\begin{minipage}[tb]{0.44\linewidth}
		\centering
		\includegraphics[width=1.8in]{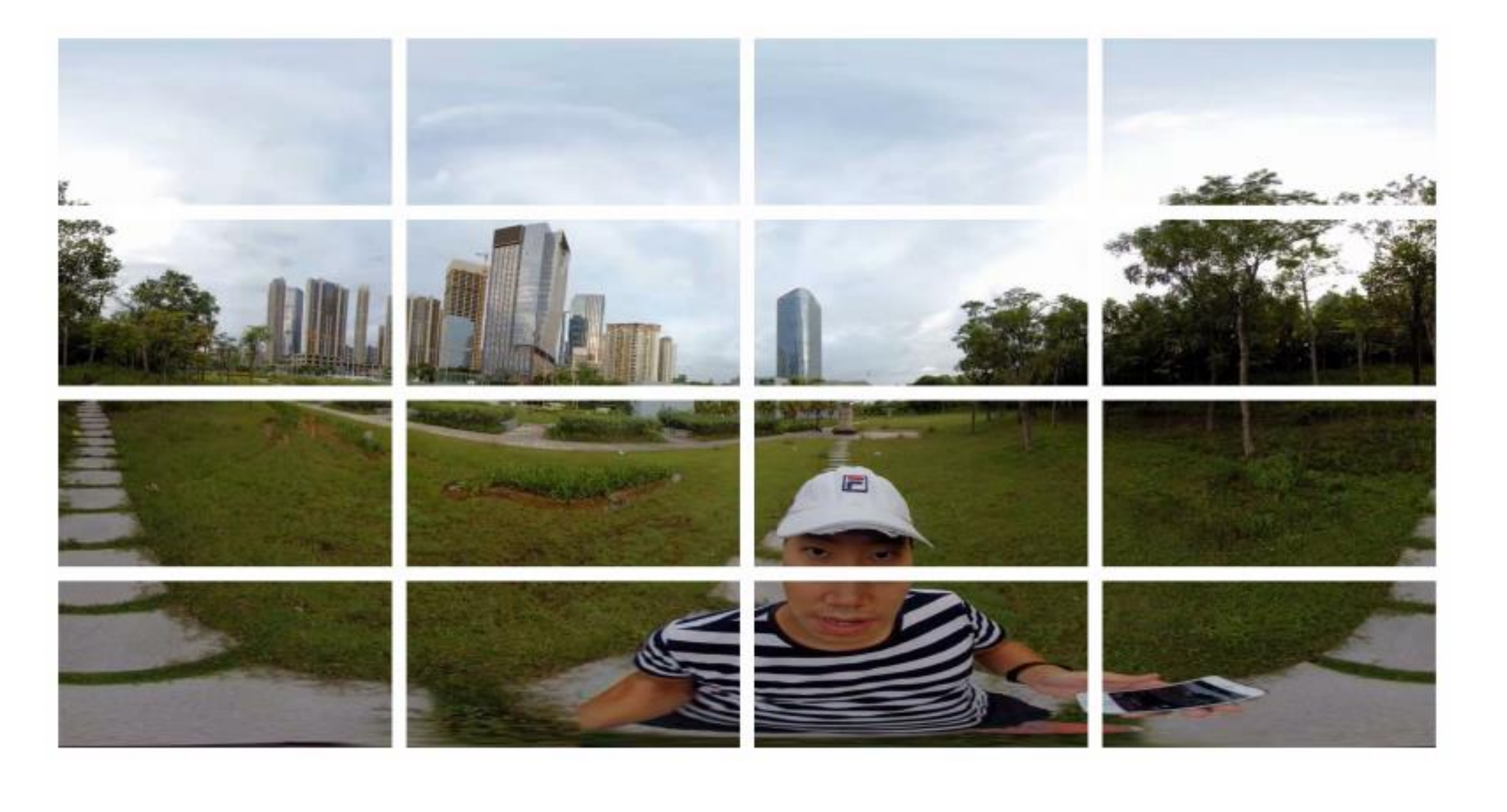}
		\caption*{(b) Tiled panoramic 360-degree video synthesized from 6 original videos.}
		\label{fig_vr_video_tile_setting}
	\end{minipage}
	\caption{Illustration of the video source}
	\label{fig:source}
\end{figure}

During the simulation, we select 6 videos captured by an Insta 360 Pro2 panoramic camera as the original video at the uplink. The highest resolution of each video is $1920 \times 1080$. The video duration is 35s, and the frame rate is 30 fps. Figure. \ref{fig:source} (a) shows a snapshot of the six original videos.
Then, we use the high efficiency video coding (HEVC) method and constant bitrate (CBR) as a bitrate control technique to compress each original video, with the bitrate representations of each video being \{1.5, 2, 2.5, 3\} Mbps. We use AVP software (Kolor Autopano Video Pro) to complete the synthesis and stitch the VR videos together. Subsequently, the panoramic VR video is divided into 16 tiles, each tile covers $90 \times 45$ degrees. We assume that the video server can provide 4 different bitrate representations for each tile at constant bitrates \{0.2, 0.6, 1, 1.4\} Mbps. Figure. \ref{fig:source}(b) shows the panoramic VR video synthesized from the 6 original videos, which is then divided into 16 tiles.
At the client side, we assume that there are 100 users wearing HTC Vive as the HDM device. The users are located at random distances from the base station, and they can request and watch 360-degree videos. The channel information can be calculated based on the fixed transmission power of the base station. During the simulation, we assume that each users' FOV ($120 \times 90$ degrees) is randomly distributed in the 360-degree video and their FOV information is transmitted to the video server in real time. 

\subsection{Simulation Results} \label{sec_simulation_results}

To better verify the performance of the proposed scheme
and make the simulation more realistic, we have conducted simulations using the real-world
network traces \cite{van2016http} with/without the perfect knowledge of future network conditions (i.e. bandwidth
prediction is 100\% correct or partially correct.). We use two baseline schemes to show the advantages of the proposed scheme. In the first method (denoted as Algorithm 2), the uplink
bandwidth resources are evenly distributed among the different cameras, and the downlink uses the
same adaptive allocation algorithm as our method. In the second method (denoted as Algorithm
3), uplink bandwidth resources are evenly distributed among the different cameras, and downlink
bandwidth resources are equally distributed to different tiles.
We first consider the case with perfect knowledge of future network condition and use the bandwidth value of each GOP as input for simulation.
Then we consider the case with imperfect knowledge of future network conditions. According to the state-of-the-art bandwidth prediction scheme \cite{1217275}, we add a Gaussian random noise, which has mean of 0 and variance of 1, to the LTE trace (with multiplying 30\% ) to simulate the prediction error.
We then perform simulation experiments using the predicted bandwidth (which differs from the real bandwidth) as the input for the optimization, and real the correct bandwidth to calculate the value of the objective function. 

The simulation results of both cases are shown in Table \ref{tab:QoE}.
From the simulation results, we can observe that the proposed algorithm can achieve a higher QoE value with all the three LTE traces. This is because we not only achieve the integration of uplink and downlink resource allocation but also efficiently allocate the resources. We can also observe that the larger bandwidth results in larger QoE value by comparing the performances with different traces. The QoE value drops due to the prediction error. This is because if the predicted bandwidth value after adding Gaussian noise is larger than the real bandwidth value, the real bandwidth may not be able to satisfy the bitrate requirement, which will cause the suspension of stalling and quality switching. If the predicted bandwidth value is less than the real bandwidth value, some bandwidth may be wasted, which will degrade the received video quality. However, the proposed algorithm still works better than the baseline scheme due to the joint consideration of uplink and downlink rate allocation.

\begin{table}[h]
		\centering
		\caption{Performance comparisons.}
		\begin{tabular}{|c|c|c|c|}\hline
			Network Trace & \multicolumn{3}{c|}{Performances}\\\hline
			& Algorithm 1 & Algorithm 2 & Algorithm 3\\\
			Bicycle Trace & 8.9057 & 7.0652 & 6.2781\\\
			Predicted Bicycle Trace & 8.0236 & 6.2537 & 5.7966\\\
			Car Trace & 6.3325 & 5.2933 & 4.7337\\\
		    Predicted Car Trace & 5.0321 & 4.1097 & 3.5282 \\\
			Bus Trace & 5.0026 & 4.0219 & 3.5537\\\
			Predicted Bus Trace & 3.9803 & 3.1102 & 2.2576\\\hline
		\end{tabular}
		\label{tab:QoE}
	\end{table}

\section{CONCLUSIONS} \label{sec_conculsion}

In this paper, we proposed a multi-user QoE-driven 360-degree video live streaming system, 
which jointly considers the uplink and downlink transmissions. In our system, the server selects the optimal bitrate settings for both the uplink and downlink channels based on the bandwidth information and the users' real-time FOV to maximize the QoE value of all users. To achieve this, we proposed an algorithm that combined the KKT condition and branch and bound method to solve the defined rate adaptation problem. Finally, the simulation results based on the real-world network traces demonstrated that our proposed algorithm outperformed other baseline schemes.

\bibliographystyle{IEEEtran}
\bibliography{joint_uplink_downlink_VR}

\begin{thebibliography}{10}
\providecommand{\url}[1]{#1}
\csname url@samestyle\endcsname
\providecommand{\newblock}{\relax}
\providecommand{\bibinfo}[2]{#2}
\providecommand{\BIBentrySTDinterwordspacing}{\spaceskip=0pt\relax}
\providecommand{\BIBentryALTinterwordstretchfactor}{4}
\providecommand{\BIBentryALTinterwordspacing}{\spaceskip=\fontdimen2\font plus
\BIBentryALTinterwordstretchfactor\fontdimen3\font minus
  \fontdimen4\font\relax}
\providecommand{\BIBforeignlanguage}[2]{{%
\expandafter\ifx\csname l@#1\endcsname\relax
\typeout{** WARNING: IEEEtran.bst: No hyphenation pattern has been}%
\typeout{** loaded for the language `#1'. Using the pattern for}%
\typeout{** the default language instead.}%
\else
\language=\csname l@#1\endcsname
\fi
#2}}
\providecommand{\BIBdecl}{\relax}
\BIBdecl

\bibitem{8647729}
J.~{Li}, R.~{Feng}, Z.~{Liu}, W.~{Sun}, and Q.~{Li}, ``Modeling qoe of virtual
  reality video transmission over wireless networks,'' in \emph{2018 IEEE
  Global Communications Conference (GLOBECOM)}, Dec 2018.

\bibitem{liu2018jet}
Z.~Liu, S.~Ishihara, Y.~Cui, Y.~Ji, and Y.~Tanaka, ``Jet: Joint source and
  channel coding for error resilient virtual reality video wireless
  transmission,'' \emph{Signal Processing}, vol. 147, pp. 154--162, 2018.

\bibitem{Liu2019}
X.~Liu, B.~Han, F.~Qian, and M.~Varvello, ``Lime: Understanding commercial 360
  degree live video streaming services,'' in \emph{Proceedings of the 10th ACM
  Multimedia Systems Conference}, ser. MMSys'19.\hskip 1em plus 0.5em minus
  0.4em\relax New York, NY, USA: ACM, 2019, pp. 154--164.

\bibitem{3123372}
X.~Corbillon, A.~Devlic, G.~Simon, and J.~Chakareski, ``Optimal set of
  360-degree videos for viewport-adaptive streaming,'' in \emph{Proceedings of
  the 25th ACM International Conference on Multimedia}, ser. MM '17.\hskip 1em
  plus 0.5em minus 0.4em\relax New York, NY, USA: ACM, 2017, pp. 943--951.

\bibitem{2017}
A.~{Ghosh}, V.~{Aggarwal}, and F.~{Qian}, ``{A Rate Adaptation Algorithm for
  Tile-based 360-degree Video Streaming},'' \emph{arXiv e-prints}, p.
  arXiv:1704.08215, Apr 2017.

\bibitem{3123291}
L.~Xie, Z.~Xu, Y.~Ban, X.~Zhang, and Z.~Guo, ``360probdash: Improving qoe of
  360 video streaming using tile-based http adaptive streaming,'' in
  \emph{Proceedings of the 25th ACM International Conference on Multimedia},
  ser. MM '17.\hskip 1em plus 0.5em minus 0.4em\relax New York, NY, USA: ACM,
  2017, pp. 315--323.

\bibitem{Konrad2017}
R.~Konrad, D.~G. Dansereau, A.~Masood, and G.~Wetzstein, ``Spinvr: Towards
  live-streaming 3d virtual reality video,'' \emph{ACM Trans. Graph.}, vol.~36,
  no.~6, pp. 209:1--209:12, Nov. 2017.

\bibitem{guo2018optimal}
C.~Guo, Y.~Cui, and Z.~Liu, ``Optimal multicast of tiled 360 vr video,''
  \emph{IEEE Wireless Communications Letters}, vol.~8, no.~1, pp. 145--148,
  2018.

\bibitem{van2016http}
J.~Van Der~Hooft, S.~Petrangeli, T.~Wauters, R.~Huysegems, P.~R. Alface,
  T.~Bostoen, and F.~De~Turck, ``Http/2-based adaptive streaming of hevc video
  over 4g/lte networks,'' \emph{IEEE Communications Letters}, vol.~20, no.~11,
  pp. 2177--2180, 2016.

\bibitem{1217275}
{Ningning Hu} and P.~{Steenkiste}, ``Evaluation and characterization of
  available bandwidth probing techniques,'' \emph{IEEE Journal on Selected
  Areas in Communications}, vol.~21, no.~6, pp. 879--894, Aug 2003.

\end{thebibliography}

\ifCLASSOPTIONcaptionsoff
\newpage
\fi

\end{document}